\documentstyle[12pt,pc96]{article}
\input epsf
\begin{document}

\title{Decay of Nuclear Giant Resonances:\\ Quantum Self-similar Fragmentation}

\author{A. Z. G\'orski\inst{1}, R. Botet\inst{2}, S. Dro\.zd\.z\inst{1}  and
M. P{\l}oszajczak\inst{1,3}}

\institute{Institute of Nuclear Physics, PL-31342 Krakow, Poland
\and Laboratoire de Physique des Solides,
B\^{a}t. 510, Universit\'{e} Paris-Sud, F-91405 Orsay, France
\and Grand Acc\'{e}l\'{e}rateur National d'Ions Lourds,
CEA/DSM - CNRS/IN2P3, BP 5027,  F-14021 Caen Cedex, France  }

\maketitle

\begin{abstract}
Scaling analysis of nuclear giant resonance transition probabilities
with increasing level of complexity in the background states is
performed. It is found that the background characteristics, typical
for chaotic systems lead to nontrivial multifractal scaling
properties.
\end{abstract}

The statistical study of spectra of quantum systems is almost as old as
quantum mechanics. In this context, the random-matrix theories (RMT)
\cite{rand}~ have been proposed as a natural theoretical framework for
describing level fluctuations in stationary systems.
The Gaussian orthogonal ensemble of random matrices, which leads to the
Wigner distribution, can describe statistical properties ( the
nearest neighbor spacing (NNS) distribution ) of nearly all time-reversal
-invariant quantum systems whose classical counterparts are chaotic
\cite{boh}~. On the other hand, the NNS distribution of levels in
quantum systems whose classical dynamics is regular
is given by the Poisson distribution. Those
'empirical' observations gave rise to 'quantum chaology' and lead
to the definition of 'chaotic', 'regular' or
'mixed' quantum systems. The fundamental limit of concepts based
on the RMT is their {\it stationary} connotation.
The nuclear giant resonance (or plasmon), which at time $t=0$~ is associated
with simple (mainly $1p$--$1h$) configurations and
fragments (decays) at later times into the complicated configurations,
 is only one of many examples of short time quantum phenomena which
necessitate the time-dependent formulation of their {\it local}
statistical properties. The effect of coupling between $1p$--$1h$ and
more complicated $np$--$nh$ configurations can be represented as a noise
or
fluctuating force acting on the observable ('macroscopic' degrees of freedom)
in the $1p$--$1h$ space. These fluctuations arise from the elimination
of the
irrelevant degrees of freedom associated with the {\it environment} of the
$1p$--$1h$ space, in favor of a small number of {\it macroscopic}
variables. In
this way one arise at the Langevin formulation of the initial multidimensional
problem in which macroscopic variables (e.g. $1p$--$1h$ collective
excitations of various quantum numbers etc.) are driven by the fluctuating
force~. In general, nature of the stochastic process depends on
the choice of the macroscopic variable. It is however worth noticing that e.g.
fluctuation properties of quantal spectra (NNS distribution : $p(x)$~)
are correctly given as the limit $t \rightarrow \infty$~ of a nonlinear
stochastic process with multiplicative fluctuations and is associated with
the Fokker-Planck equation\cite{sch}~:

\begin{equation}
\label{f3}
\frac{\partial}{\partial t}p(x,t) = -\frac{\partial}{\partial x}
[(Dx - Bx^{1+\gamma} + \frac{1}{2}Qx)p(x,t)] + \frac{Q}{2}
\frac{{\partial}^{2}}{\partial x^{2}} (x^{2}p(x,t))~~~~\ ,
\end{equation}
where parameters $B,D,\gamma$~ can be identified with parameters of the
fundamental Hamiltonian of the system, and $\gamma \equiv 2D/Q$~
equals $1$~ or $2$~ for Poisson or Wigner distributions respectively.
In between those two limiting cases, i.e. for
$ 1 < \gamma < 2$~, one obtains from (\ref{f3})
the Brody distribution\cite{brod}~:
$p(x) = Nx^{\gamma - 1} \exp(-{\tilde b}x^{\gamma})$~ with
${\tilde b} \equiv B/D$~, which describes well the NNS in
mixed quantum systems \cite{shr}~. This stochastic process (\ref{f3})
resembles the binary, self-similar and conservative random fragmentation
process \cite{bp}~ which is known to		
yield 'universal' behaviors, independently of the precise fragmentation
mechanism \cite{bp1}~. Generality of this stochastic process
 makes plausible the hypothesis of universality also in the
 strength fragmentation associated with the giant resonance decay.
This work reports about the investigation of this intriguing quantum
phenomenon.

The generic statistical framework for the description of
fragmentation aspects of the giant resonance decay is provided by the Nakajima
- Zwanzig rate equations~:
\begin{eqnarray}
\label{f4}
i\frac{\partial}{\partial t} f_1(t) - \sum_{1^{'}}^{} H_{11^{'}} f_{1^{'}}(t) =
\sum_{1^{'}}^{} \int_{0}^{t} d\tau f_{1^{'}}(t-\tau ) v_{11^{'}}(\tau )~~~~~\ ,
\end{eqnarray}
where the non-local collision term~:
\begin{eqnarray}
\label{f5}
v_{11^{'}}(\tau ) = -i\sum_{2}^{}H_{12}H_{21^{'}} \exp (-iH_{22}\tau )
\end{eqnarray}
is expressed in terms of the matrix elements of the Hamiltonian in the
$1p$--$1h$ ($\mid 1 >$~) and $2p$--$2h$ ($\mid 2 >$~) background
spaces \cite{dnsw}.
$f_{1^{'}}$~ in  (\ref{f4}) are the time-dependent probability
amplitudes of the macroscopic, collective state in the space $\mid 1 >$~.

\begin{figure}[bht]
\centering
\epsfxsize=10.0cm
\mbox{
\epsfbox{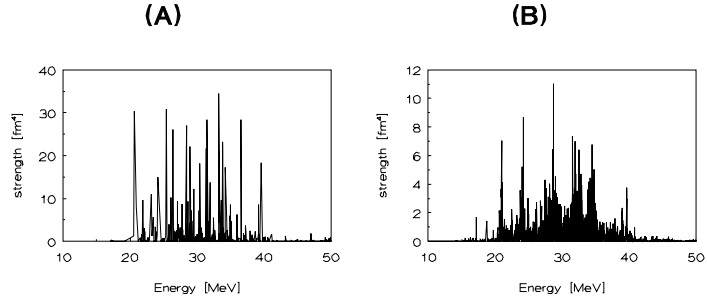}
}
\caption{Isovector strength distribution for the regular (A) and
chaotic (B) case.}
\label{Fig1}
\end{figure}


 In the limit $t \to \infty$ this has been studied recently
\cite{dnw}~  for the
giant quadrupole resonances of $^{40}Ca$ in the truncated subspace of
$1p$--$1h$ (26 states) and $2p$--$2h$ (3014 states) excitations.
Fig. 1 shows the resulting isovector quadrupole strength distribution
$\rho_i$ ($\rho_i = \mid < 0 \mid \hat f \mid i > \mid^2$,
$\hat f$ is the one--body operator and $\mid i >$ diagonalizes
$\hat H$ in $\mid 1 > \oplus \mid 2 >$)
in the two model cases: (A) no residual interaction in $2p$--$2h$
subspace which leads to the Poisson level fluctuations and can be
associated with regular dynamics and (B) full residual interaction
included which results in Wigner fluctuations and thus can be
interpreted in terms of chaotic dynamics \cite{dnsw}~.
Significant redistribution of the strength is observed  when going
from (A) to (B).
The most interesting effect is more uniform distribution, even
resembling a certain kind of self--similar structure regarding
the clustering and the relative size of the transitions.
This points to the need of more systematic, multifractal analysis.

\begin{figure}[bht]
\centering
\epsfxsize=12.0cm
\mbox{
\epsfbox{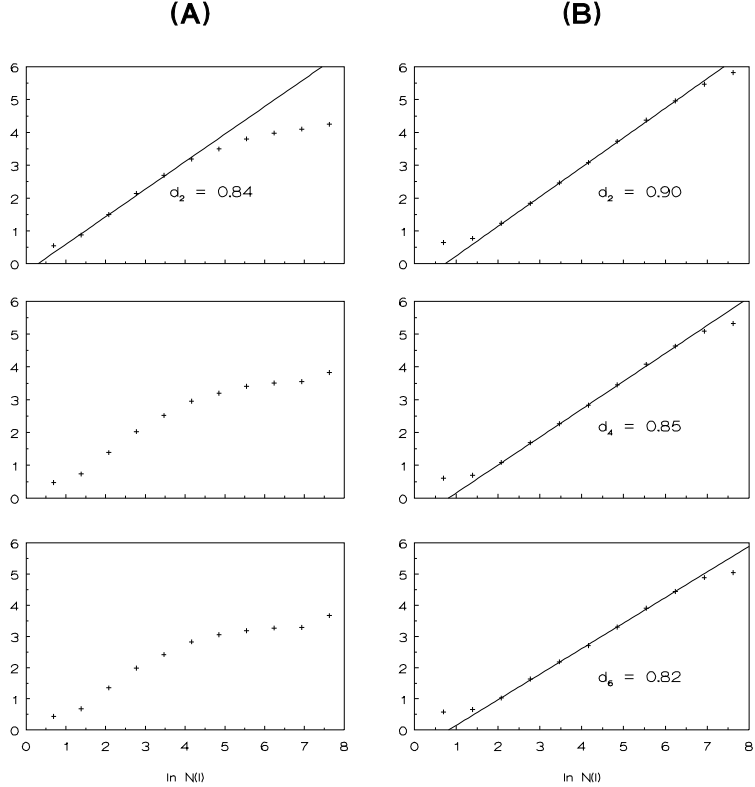}
}
\caption{Energy levels weighted by transition probabilities scaling
in the regular (A) and chaotic case (B) for $q=2, 4, 6$. }
\label{Fig2}
\end{figure}


 To this end
we have investigated the sets of {\it pairs} of points: the energies
and the transition probabilities ($\{ E_i, \rho_i\}$).
The energy spectrum of a typical quantum systems has the fractal
dimension equal to one.
However, taking into account probabilities $\rho_i$ we have more
complicated structure. Because to each energy corresponds only one
probability the whole structure should have dimension in the range:
$0 \le d_q \le 1$.

 Therefore, we shall treat the value $\rho_i$ as a number of "points" in
the corresponding ($i$--th) box. Hence, the fractal structure of the
set $\{ E_i \}$ will be modified by the transition probabilities.
To have measure with
the proper normalization we define our measure $P_i(l)$ as:

\begin{equation}
\label{OURMEASURE}
p_i(l) =  P_i(l) \equiv
\left[ \ \sum_{\rm all\ }E_i \ \rho_i \ \right]^{-1} \ \times
\sum_{E_i\in\ {\rm i-th\ box}} \, \rho_i
\ ,
\end{equation}

\noindent where in the numerator summation goes over probabilities
whose energies are included in the $i$-th box.
Here again the measure $P_i(l)$ is properly normalized:
$\sum_i P_i(l) = 1$.
The scaling exponent ("fractal dimension") with the measure
(\ref{OURMEASURE})
we will denote by $D_q$, while $d_q$ we reserve for the standard
fractal dimension.
From (\ref{OURMEASURE})
it is clear that for $q=0$ (capacity dimension)
probabilities $\rho_i$ do not contribute to $\chi_0(l)$,
as $M(l)$ is the number of boxes with {\it any} non--zero number
of data points (energies in our case).
Hence, we have: $D_0 = d_0$, the last being the standard capacity
dimension of the energy spectrum.

 The input data consist of the order $\sim 2^{11}$ data points,
the number sufficient
to display exponential scaling but one should have in mind that
some statistical errors will be present, as the fractal dimension
formula contains the $l\to0$ (or, equivalently, the $n\to\infty$) limit.

 In fact, for the chaotic case we have got a fairly good scaling
in the range of about 8 points in the log--log plot
as can be seen from Figs. 2(B).
This nice scaling has been considerably worsened in the regular case
(Fig. 2(A)~).
In the special case of the capacity dimension we get $D_0 \simeq 1$,
as in this case the scaling exponents are determined solely by the
energy distribution and this has not been plotted in Fig. 2(B).
 The regular case is also plotted for comparison, even though
scaling is very poor and choice of the scaling exponents
is difficult in this case. Hence, for $q>2$ the linear fits have not
been plotted on Figs. 2(A).

Further studies are still needed to disclose salient features of
short time quantal phenomena in the framework of the conservative
and self-similar random fragmentation process \cite{bp}~, which
in the asymptotic limit
$t \rightarrow \infty$~ provides a good description of the
{\it local} statistical properties of quantum systems both in their
chaotic and regular limits as well as in the mixed limit.
The ubiquity of this fragmentation process \cite{bp}~, both at small
(microscopic) and large (macroscopic) scales, makes possible a new
insight into the relation between chaoticity or regularity
of classical systems and the corresponding statistical properties of the
associated quantum systems.
\noindent
\paragraph{Acknowledgments.\/}
This research was partially supported by
grant KBN 2 P03B 140 10.

\end{document}